\documentclass[english]{llncs}
\usepackage[T1]{fontenc}
\usepackage[latin9]{inputenc}
\usepackage{babel}
\usepackage{float}
\usepackage{url}
\usepackage{graphicx}
\usepackage[unicode=true,pdfusetitle,
 bookmarks=true,bookmarksnumbered=false,bookmarksopen=false,
 breaklinks=false,pdfborder={0 0 0},pdfborderstyle={},backref=false,colorlinks=false]
 {hyperref}

\makeatletter

\floatstyle{ruled}
\newfloat{algorithm}{tbp}{loa}
\providecommand{\algorithmname}{Algorithm}
\floatname{algorithm}{\protect\algorithmname}

\usepackage{bussproofs}
\floatname{algorithm}{Listing}

\makeatother

\usepackage{listings}
\begin{document}

\title{FMS: Functional Programming as a Modelling Language}

\author{Ingmar Dasseville, Gerda Janssens}

\institute{KU Leuven, Dept. of Computer Science, B-3001 Leuven, Belgium.\\
\texttt{ingmar.dasseville@cs.kuleuven.be}~\\
\texttt{gerda.janssens@cs.kuleuven.be}}
\maketitle
\begin{abstract}
In this paper we introduce the Functional Modelling System (FMS).
The system introduces the Functional Modelling Language (FML), which
is a modelling language for NP-complete search problems based on concepts
of functional programming. Internally, we translate FML specifications
to an Answer Set Program to obtain models. We give a general overview
of the new FML language, and how this language is handled in the system.
We give a step-by-step walkthrough of the system, pointing out what
features are in place, and what improvements are still possible.
\end{abstract}

\section{Introduction}

We have published a theoretical framework in which we explain how
lambda calculus could be used to define higher order logics\cite{iclp/DassevilleHBJD16}.
Based on these ideas we now present a practical Functional Modelling
System (FMS), powered under the hood by an Answer Set Programming
(ASP) solver. The system is focused on solving NP-complete search
problems, the same class of problems its underlying engine - ASP -
can solve. In this paper we give a high level overview of the system,
point out which techniques and algorithms are already implemented
and highlight the places where there is low-hanging fruit for improving
the system. In Section \ref{sec:Language} we give an overview of
the language of the system, both on the external and the internal
level. In Section \ref{sec:General-Workflow} we explain what steps
a file goes through to produce an answer. In Section \ref{sec:Optimisations}
we take a deeper look into what kind of optimisation techniques are
present in the system. We conclude in Section \ref{sec:Implementation}
with information on the tooling and availability of the system. 

\subsection{Context}

In this paper we use the concept of \emph{modelling language }to denote
a language which is used to express a set of rules to which solutions
should adhere. Typically, modelling languages allow for naturally
expressing NP-complete problems. A solution for this set of rules
is called a model. Typical examples of such languages are Minizinc\cite{conf/cp/NethercoteSBBDT07},
ASP\cite{asp/2001} and FO(.)\cite{iclp/Blockeeletall12}, the language
of IDP3. The new Functional Modelling Language (FML) is an addition
to this set of modelling languages. Most previous sytems have their
roots in logic programming, and first order logic. The new FML language
is based on constructs which are traditionally only found in functional
programming languages. With these constructs it is still possible
to use traditional first order logic specifications, but there are
also new modelling possibilities. 

The only other fully higher order modelling language that the authors
are aware of is ProB\cite{LeuschelButler:FME03}. The expressive power
of ProB is very high but the solving capabilities of this language
are limited. The system is mainly used to model dynamic systems and
not to solve search problems.

Monadiccp\cite{DBLP:journals/jfp/SchrijversSW09} and ersatz are Haskell
libraries, through which it is also possible to define search problems
using a functional language. It allows the user to define constraint
satisfaction problems using Haskell, which is then translated to respectively
a constraint programming or a SAT solver. While the host language
is a fully higher-order language, the problems that can be specified
in monadiccp are only first order. Ersatz does have the notion of
relations, but somebody writing the specification has to think about
the possible range of values while declaring the relation. FMS takes
this one step further as the domain of the function is derived from
the specification.

\subsection{Search Technologies}

There are a multitude of existing technologies apt for solving search
problems. SAT solvers have become very efficient and are the standard
for a number of problems, but there are also a lot of other technologies:
SMT (SAT modulo theories) which extends SAT with higher level concepts,
constraint programming systems, mixed integer programming, \ldots.
FMS does not contribute to these search technologies, but to the modelling
techniques for these.

For FMS, ASP was chosen. ASP solvers use a SAT solver with some extensions
for recursive logic definitions. ASP was chosen because the language
itself is already relatively high level, so it is easier to build
another language on top of ASP compared to other paradigms such as
SAT.

\subsection{First Order or Higher Order?}

FML supports higher order expressions: functions which take functions
as input and return new functions. Higher order sets are also supported:
it is possible to define \lstinline!s! as the set $\left\{ \left\{ 1,4\right\} ,\left\{ 1,6,7\right\} ,\left\{ 5,4,9\right\} \right\} $. 

We have a polynomial translation to ASP which has first order semantics,
from this it follows that the semantics of our language is not higher
order. The higher order aspects of the FML language are syntactical
sugar for a first order encoding of higher order functionality.

Nevertheless, this approach allows for modelling techniques which
were not possible before in modelling languages. For instance in Listing
\ref{alg:Higher-Order-Theories} you can see an FML specification
which declares the constants \lstinline!c! and \lstinline!d! as
numbers between 3 and 5. The next line defines the set \lstinline!s!
as the set containing the doubling function, the function which adds
\lstinline!c! to a number and the function which multiplies a number
by \lstinline!c!. The last line starts with \lstinline*!* which
means ``forall''. So the line means: for every function \lstinline!f!
in the set \lstinline!s!, \lstinline!f! applied to \lstinline!d!
is smaller than \lstinline!10!. There is exactly one model for this,
which maps both \lstinline!c! and \lstinline!d! to $3$.

\begin{algorithm}
\begin{lstlisting}
c :: element of {3..5}.
d :: element of {3..5}.

s := {\x -> x * 2, \x -> x + c, \x -> x * c}.

! s (\f -> f d < 10).
\end{lstlisting}

\caption{\label{alg:Higher-Order-Theories}Quantification of functions (in
the Full language)}
\end{algorithm}

\subsection{Advantages}

There are a number of advantages when comparing FML to existing modelling
languages. One driving factor for higher order language is the easy
abstraction of concepts. Suppose that you want to want to find 2 different
solutions of the N-queens problem so that no two queens occur in the
same place. In traditional ASP or MiniZinc (a constraint programming
language), one would need to write down the N-queens constraints twice:
once for each of the two solutions. Extensions for ASP have been introduced
to introduce templates\cite{DBLP:conf/nmr/IanniIPSC04}, which circumvent
this problem but make a fundamental difference between first order
and higher order predicates. FML does not require the user to differentiate
between first and higher order terms and allows to easily reuse the
higher order predicate \lstinline!nqueens! indicating that \lstinline!solution!
is a solution to the nqueens problem. The FMS modelling for this problem
can be seen in Listing \ref{alg:Nqueens}.

A second advantage of the FML is the propagation of the domains. In
most traditional modelling languages, whenever you declare a function
or predicate, you need to specify a finite domain. This would disallow
a definition of prime numbers as short as ``\lstinline*prime x := ! {2..x-1} (\y -> x 
(x is prime if for all numbers y between 2 and x-1, x modulo y is
larger than 0) because this is a function with an infinite domain
(all integers). FML allows this by automatically deriving the relevant
domain of the function from the constraints. 

\begin{algorithm}
\begin{lstlisting}
domain := {1..8}.

solution1/1 :: function to domain.
solution2/1 :: function to domain.

alldiff solution f := ! domain (\x -> 
                        ! domain (\y -> x ~= y => f x ~= f y)).

nqueens solution := ! {
                        solution,
                        \x -> x - solution x,
                        \x -> x + solution x
                      } (alldiff solution).
nqueens solution1.
nqueens solution2.
! domain (\x -> solution1 x ~= solution2 x).
\end{lstlisting}

\caption{\label{alg:Nqueens}Find 2 N-queen solutions which have no shared
queens}
\end{algorithm}

\section{Language\label{sec:Language}}

We differentiate between the \emph{Core} language and the \emph{Full}
language of FMS. We start by introducing the Core language in Section
\ref{subsec:Core}. The Core language represents the essentials of
FML. The Full language extends this with syntactical sugar for ease
of use and is introduced in Section \ref{subsec:Full-Language}. Internally
in the system, all expressions are represented using the Core language.
It makes internal transformations easier than a representation that
incorporates all details of the Full language.

\subsection{Core Language \label{subsec:Core}}

FML is based on the lambda calculus. This lambda calculus is extended
with some common constructs (such as let-bindings and numbers) and
less common constructs (such as the built-in notion of a set). There
are ten language constructs in the Core language. Nine with a semantic
meaning, and one to track the values which are needed in the output.
This language does not have an explicit syntax as it corresponds to
the internal representation of the language. However, it can be useful
to see an approximate BNF for this Core language which can be seen
in Listing \ref{alg:Core-BNF}. The language constructs you can see
in the BNF are:

\begin{algorithm}
\begin{lstlisting}
e := x                          (variable)
   | e e                        (application)
   | \x -> e                    (lambda abstraction)
   | let (x := e)* in e         (let binding)
   | i                          (injection)
   | case e of (pattern -> e)*  (case expression)
   | b                          (builtin)
   | {e*}                       (set expression)
   | h(e*)                      (herbrand expression)
   | outputexp(s, e)            (output expression)

pattern := _                    (don't-care pattern)
         | h[pattern*]          (constructor match)
         | x                    (variable match)
\end{lstlisting}

\caption{\label{alg:Core-BNF}BNF for the Core language}
\end{algorithm}

\begin{description}
\item [{Variable}] A traditional reference to a bound identifier
\item [{Application}] This construct applies a function to an argument.
\item [{Lambda}] Internally, this is the only way a function can be declared.
Lambdas abstract away one single variable. If multiple arguments are
required, this language construct should be nested. Note that a tuple
counts as a single argument.
\item [{Let~binding}] Introduces some local definitions in a new expression.
A binding binds a variable to an expression, possibly recursively.
This is the only way to bind a term recursively.
\item [{Injection}] This is a way to inject an arbitrary JVM object into
the language. In the current system, it is only used to inject integers
and strings. In the future this construct could be used to reason
over arbitrary JVM objects.
\item [{Case}] This language construct handles all pattern matching. This
can be seen as a generalised form of an if-then-else construct. An
if-then-else construct would only handle patterns: true and false.
\item [{Builtin~symbol}] There is quite a list of builtin functions for
the FMS. The reason for this is not always that we can't define them
in the language, but that they are common functions and a dedicated
method for translating them leads to a more efficient translation
to ASP. Builtin functions include arithmetic operations (\lstinline!+!,\lstinline!*!),
boolean functions (\lstinline!&!,\lstinline!|!) and set functions
(forall: \lstinline*!*, exists: \lstinline!?!).
\item [{Set~Expression}] Sets could be emulated through the use of constructors
but for efficiency reasons we opted to have sets as an explicit built-in
datatype of the language. This language construct makes a finite set
containing some expressions. The type system enforces that all elements
of a set are of the same type. To represent all set expressions contained
in the Full language, this Core construct needs to be combined with
some builtin symbols.
\item [{Herbrand~Expression}] This construct corresponds to the declared
constructors. It is a named constructor optionally applied to some
arguments. These constructors can be pattern matched using the Case
construct.
\item [{Output~Expression}] \begin{sloppypar} This language construct
is the only one which says nothing about the actual meaning of the
expression but is a construct to annotate that the expression it contains
should be outputted using a certain name. \lstinline!4 + OutputExp("a",5)!
would denote an expression that evaluates to 9, but the subexpression
\lstinline!5! should be outputted under the name \lstinline!"a"!.
Most languages use some kind of print statements for this. But as
we don't have those in a modelling language like FML, we have to track
the values we want to print in another way.  \end{sloppypar} 
\end{description}

\subsubsection{Types}

FML is strongly typed, this means that every expression has a type
which is enforced at compile time, before trying to run the program.
The language is also implicitly typed, this means that these strong
types are not necessary parts of the syntax and it is possible that
the input file contains no type annotations. 

The type system itself is a Hindley-Milner type system, which includes
polymorphism through universal quantification over types. E.g. The
forall quantifier has the following type: \lstinline!$\forall$ a. Set a -> (a -> Bool) -> Bool!.
In the implementation itself, the universal quantification is left
implicit.

\subsubsection{Constructors}

It is possible to declare new constructors and deconstruct values
which are made of these constructors using the case-construct. An
example using peano arithmetic is shown in Listing \ref{alg:Constructor-Example}.

\begin{algorithm}
\begin{lstlisting}
s/1 :: constructor.
nil/0 :: constructor.

minusOne x := case x of
		s [ a ] -> a;
		nil []    -> nil;.
\end{lstlisting}

\caption{\label{alg:Constructor-Example}Minus One in Peano Arithmetic (in
the Full language)}
\end{algorithm}

\subsection{Full Language\label{subsec:Full-Language}}

The Full language extends the Core language with a lot of syntactic
sugar. The process of converting the Full language to the Core language
is called desugaring. We will take the graph coloring problem as a
leading example throughout this section. The formulation of a graph
coloring problem in FML can be seen in Listing \ref{alg:Graph-Coloring}. 

In the Full language, we discern three different kinds of statements:
declarations, definitions and constraints. Declarations introduce
new symbols for which the interpretation is not given through a definition
(e.g. symbol colorof). Definitions introduce new symbols of which
the interpretation is fixed (e.g. the border relation). Constraints
are boolean expressions which must be true in every model (e.g. adjacent
nodes have different colors).

\begin{algorithm}
\begin{lstlisting}
//Definitions of given sets
borders := {("a","b"), ("b","c"), ("c","a")}.
colors := {1..3}.

//Declaration of the interpretation we are looking for
colorof/1 :: function to colors.

//Constraint: For all borders (x,y) the color of x 
//            should be different than that of y
! borders (\(x,y) -> colorof x ~= colorof y).
\end{lstlisting}

\caption{\label{alg:Graph-Coloring}Graph Coloring (in the Full language)}
\end{algorithm}

\subsubsection{Declarations}

Declarations have the form: \lstinline!name :: declarationkind.!.
The most important declarations are:
\begin{itemize}
\item \lstinline!e :: element of set! where \lstinline!set! is an expression
evaluating to a set. The interpretation of \lstinline!e! will be
a single element of the interpretation of the set.
\item \lstinline!s :: subset of set! where \lstinline!set! is an expression
evaluating to a set. The interpretation of \lstinline!s! will be
a subset of the interpretation of the set.
\item \lstinline!f/arity :: function to set! where \lstinline!set! is
an expression evaluating to a set and \lstinline!arity! a natural
number. The interpretation of \lstinline!f! will be a function of
arity \lstinline!arity! with codomain \lstinline!set!.
\item \lstinline!c/arity :: constructor! where \lstinline!arity! is a
natural number. The interpretation of \lstinline!c! will be a new
constructor function of arity \lstinline!arity!.
\end{itemize}
\begin{sloppypar}There are also extra declarations like \lstinline!proposition!
or \lstinline!predicate!. These can be rewritten in terms of other
declarations such as \lstinline!element of {true,false}! or \lstinline!function to {true,false}!\end{sloppypar}
\begin{example}
A graph coloring problem needs one declaration: the coloring of the
nodes, which is a function to colors.
\end{example}

\subsubsection{Definitions/Constraints}

\begin{sloppypar}A definition has the form \lstinline!identifier := expression.!.
A constraint is just an expression by itself. Definitions can be recursive
and mutually dependent (e.g. mutually dependent odd and even definitions).
There is builtin support for first order logic-connectives, quantifiers,
arithmetic and set-expressions. The quantifiers forall (\lstinline*!*)
and exists (\lstinline!?!) are slightly different from the traditional
notation in first order logic. They are higher order functions with
two arguments: the set they are quantifying over and a boolean function
explicitating the property which needs to hold.\end{sloppypar}
\begin{example}
A graph coloring problem needs two defined relations: the borders
and the colors. These can just be given as a set expression. There
is one constraint, which quantifies over all borders. We can deconstruct
the tuples in the border-relation in place using \lstinline!\(x,y) -> $\ldots$!.
The body of the forall quantifier just states that the colors x and
y should be different.
\end{example}

\subsubsection{Set Expressions}

Set expressions are a language feature based on the set-builder notation
of mathematics. The set expressions in the Full language are much
more general than those of the Core language. The syntax is heavily
inspired by the notation in Haskell and Python. Set expressions can
be replaced by a series of map and filter operations on sets but allow
for a more readable syntax. Point-wise application of a function f
over a set s looks like: \lstinline!{f x || x <- s}!. Which can be
read as: the set of evaluation of \lstinline!f x! where \lstinline!x!
is in the set \lstinline!s!. Or the subset of s for which the predicate
p holds can be written as: \lstinline!{x || x <- s, p x}!.

\section{General Workflow\label{sec:General-Workflow}}

\begin{figure}
\begin{centering}
\includegraphics[clip,width=0.68\columnwidth]{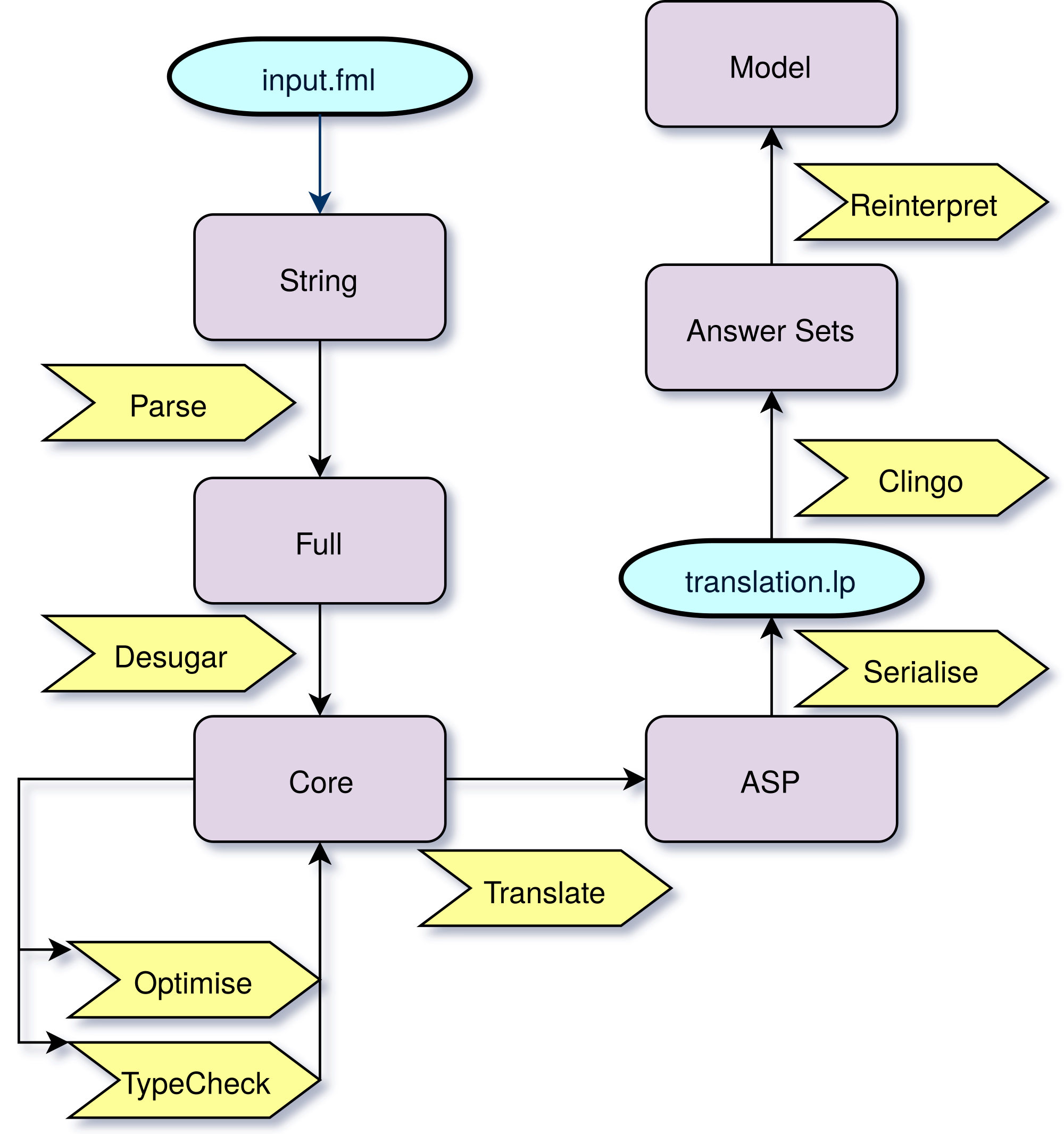}
\par\end{centering}
\caption{\label{fig:The-general-workflow}The general workflow of FMS}
\end{figure}

A lot of inspiration for the general workflow of FMS came from GHC\cite{GHC}.
FMS can be seen as a compiler to Answer Set Programming and most of
the steps in the compiling process of GHC is relevant to FMS. An overview
of the workflow can be seen in Figure \ref{fig:The-general-workflow}.
In this section we go over these steps to investigate them in more
detail, from the file containing an FML specification to the models.

\subsection{Parse and Desugar\label{subsec:Parsing-and-Desugaring}}

Parsing an FML specification is done using the ANTLR4\cite{parr2013definitive}
system. Based on a BNF grammar, ANTLR4 generates an Abstract Syntax
Tree (AST) of an FML specification. This syntax tree is then desugared
into the Core language, as defined in Section \ref{subsec:Core}.
Checking whether any identifiers are used out of scope is considered
the last step in the desugaring process. Once a string is fully parsed
and desugared, it becomes a self-contained well-formed Core expression
without free variables.
\begin{example}
The expression \lstinline!f 1 := 0. f x := 1! 

is desugared to \lstinline!f := (\y -> case y of 1 -> 0 ; x -> 1)!
\end{example}

\begin{example}
The expression \lstinline!(a,b) := f 5! is desugared to 

\begin{lstlisting}
temp := f 5. 
a := case temp of (a,_) -> a. 
b := case temp of (_,b) -> b.
\end{lstlisting}

\noindent The reasoning behind this desugaring is that all pattern
matching needs to be done in the case-construct and definitions can
internally only define a single variable, but the unsugared line defined
\lstinline!a! and \lstinline!b! using a single assignment.
\end{example}

Desugaring complex set expressions results in usages of the bind builtin.
The bind operation is the same as the bind for monads in Haskell\cite{DBLP:journals/mscs/Wadler92}
sometimes known as flatmap. Flatmap (specialised for sets) boils down
to a pointwise application followed by a flattening. Desugaring sets
could potentially be further improved by using an applicative style
instead of a monadic style wherever possible\cite{marlow2016desugaring}.
An applicative style omits the flattening part and can be more efficient,
but has less expressivity.

\begin{example}
The flattening of a set \lstinline!ss! can be expressed by the following
expression: \lstinline!{x || s <- ss, x <- s}!

Which is desugared to: \lstinline!bind ss (\s -> bind s (\x -> {x}))!
\end{example}

Our approach directly transforms a FML specification to the Core language,
which has some disadvantages. While the location information of symbols
is stored inside the Core language data structures, some details about
the original representation get lost. This means that every error
which is detected in the file after desugaring is less clear than
if the analysis would have been done on the original representation.
This mainly concerns errors about out-of-scope identifiers and typing
errors, these are exactly the ones which are checked on the desugared
version in GHC\cite{GHC}. 

\subsection{Optimise}

We could directly translate the Core language to ASP. However, there
are many specification transformations which make the specification
simpler. For this reason a number of optimisation techniques are applied
to the expressions in the Core language. These optimisations deserve
their own section and are further described in Section \ref{sec:Optimisations}.

\subsection{Type Check}

Types have no semantic value in FML. This means that the evaluation
of a specification can be done without taking typing information into
account. Nevertheless, types are an important tool to detect mistakes
in a specification, both for the end-user who writes a specification,
and for a compiler writer when implementing optimisations to check
that the optimised version of the specification is still well-typed.
A fairly standard Hindley-Milner\cite{MIT/Pierce2002} type system
is implemented on the Core language (and not on the Full language).
As explained in Section \ref{subsec:Parsing-and-Desugaring}, a second
implementation on the Full language would improve the error messages
produced by the type system. Note that doing type checks on the Full
language does not fully eliminate the need for a type checker at the
Core level, type checking on Core is still useful to detect errors
in the optimisers. The error diagnosis in case of a badly typed expression
is also very vague in the current state of the system. Many extensions
of the standard algorithm exist which would lead to clearer error
messages\cite{wazny2006type}, a future version of the system would
benefit from those.

\begin{algorithm}
\begin{lstlisting}[numbers=left]
:-not bool(X),result(X).
member(s0,X0):-X0=1..3.
out("colors",s0).
{lamInter(l0,X2,X1):member(s0,X1)}==1:-lamDom(l0,X2).
out("colorof",l0).
member(s1,("a","b")).
member(s1,("b","c")).
member(s1,("c","a")).
out("borders",s1).
lamDom(l0,X4):-member(s1,X3),(X4,X5)=X3.
lamDom(l0,X5):-member(s1,X3),(X4,X5)=X3.
bool((b0,X3)):-X6<>X7,lamInter(l0,X4,X6),
		member(s1,X3),(X4,X5)=X3,lamInter(l0,X5,X7).
bool(b1):-not bool((b0,X3)),member(s1,X3).
bool(b2):-not bool(b1).
result(b2).
\end{lstlisting}

\caption{\label{alg:Graph-ColoringASP}Translation of Listing \ref{alg:Graph-Coloring}
to ASP}
\end{algorithm}

\subsection{Translate}

The translation is arguably the most important part of the system.
This part of the system does the job of translating the Core language
to an Abstract Syntax Tree representation of ASP. Although the algorithm
is complex, it is only a minor part of the actual code. The inner
workings of the translation process is considered out of scope for
this paper, but we give some intuitions using Listing \ref{alg:Graph-ColoringASP},
which is the translation of the graph coloring example of Listing
\ref{alg:Graph-Coloring}. \footnote{The translation presented is slightly simplified for readability.
The only difference lies in the omission of superfluous empty tuples} More details about translating a functional language into ASP is
given in another paper presented at WFLP 2018.

The ASP programs that FMS produces contain three kinds of statements.
Declarations are translated to choice rules. Definitions and constraints
are translated to definitions and there is one constraint: it enforces
that the evaluation of the constraints is true. This constraint can
be seen on line 1. Sets are characterised by a name such as \lstinline!s0!,
and \lstinline!member/2! represents the membership relation between
sets and their elements. Line 2 in the translation defines the set
\lstinline!s0! as the numbers of 1 to 3. In line 3 you can see that
this set represents the set of colors of the input. The \lstinline!lamInter/3!
relation is used to define the relation between a function (the first
argument), an input (the second argument) and its output (the third
argument). Line 4 defines \lstinline!l0! as an uninterpreted function
mapping every member of \lstinline!lamDom! (the predicate which is
used to represent the relevant domain of the function) to a member
of the set \lstinline!s0!. This corresponds to the colorof function,
this is explicitated in line 5. Lines 6-9 do the same thing for borders
as lines 2-3 did for colors. Lines 10 and 11 define that the relevant
domain are the first and second components of the elements in the
set of borders. Line 12-15 make up the translation of the constraint.
Just like functions, booleans are given a name like \lstinline!b0!.
The \lstinline!bool/1! relation is used to indicate whether the boolean
is considered true. In line 16, it can be seen that the boolean \lstinline!b2!
is considered the top constraint.

As a whole this ASP program has models which correspond to the 6 possible
colorings of the graph. One example answer set can be seen in Listing
\ref{alg:Graph-ColoringResult}. Once the translation is obtained,
we send this ASP program further into our workflow towards Clingo.

\begin{algorithm}
\noindent\begin{minipage}[t]{1\columnwidth}%
\begin{minipage}[t]{0.45\columnwidth}%
\begin{lstlisting}[firstline=1,lastline=10,breaklines=true]
result(b2) 
bool((b0,("a","b"))) 
bool((b0,("b","c"))) 
bool((b0,("c","a")))
bool(b2) 
lamDom(l0,"a") 
lamDom(l0,"b") 
lamDom(l0,"c") 
lamInter(l0,"a",1) 
lamInter(l0,"b",3) 
lamInter(l0,"c",2) 
member(s0,1) 
member(s0,2) 
member(s0,3) 
member(s1,("a","b")) 
member(s1,("b","c")) 
member(s1,("c","a")) 
out("borders",s1) 
out("colorof",l0) 
out("colors",s0) 
\end{lstlisting}
\end{minipage}%
\begin{minipage}[t]{0.45\columnwidth}%
\begin{lstlisting}[firstline=11,lastline=20,breaklines=true,firstnumber=11]
result(b2) 
bool((b0,("a","b"))) 
bool((b0,("b","c"))) 
bool((b0,("c","a")))
bool(b2) 
lamDom(l0,"a") 
lamDom(l0,"b") 
lamDom(l0,"c") 
lamInter(l0,"a",1) 
lamInter(l0,"b",3) 
lamInter(l0,"c",2) 
member(s0,1) 
member(s0,2) 
member(s0,3) 
member(s1,("a","b")) 
member(s1,("b","c")) 
member(s1,("c","a")) 
out("borders",s1) 
out("colorof",l0) 
out("colors",s0) 
\end{lstlisting}
\end{minipage}%
\end{minipage}

\caption{\label{alg:Graph-ColoringResult}An Answer Set of the program in Listing
\ref{alg:Graph-ColoringASP}}
\end{algorithm}

\subsection{Clingo}

We chose the ASP solver Clingo as a backend for our system. We communicate
with Clingo via text files and use JSON output facilities of Clingo
to be able to easily parse the answer sets. From the viewpoint of
implementing FMS this is a very easy step. We print the ASP AST to
a text file, call Clingo and reparse its output. The advantages of
file-based communication are that it is easily debugged as it is human
readable and it is easy to replace Clingo with another solver using
the ASP Core\cite{aspcore2} standard. It would be more efficient
to directly use the internal APIs of Clingo to build the ASP program.
Clingo produces answer sets, which represent models for the problem,
but it is still not the end of our workflow.

\subsection{Reinterpreting the Answer Sets}

Answer sets map one-to-one to FML models. However, it is not obvious
how the answer sets from Clingo relate to the original problem statement,
as can be seen in Listing \ref{alg:Graph-ColoringResult}. The answer
set possibly contains higher order functions and sets, which are relatively
complicated when encoded in answer set. This step reorders the information
of the answer set into a more natural format. 
\begin{example}
The answer set in Listing \ref{alg:Graph-ColoringResult} is transformed
to the following model for the end user:

\begin{lstlisting}
{borders=[(a,b), (b,c), (c,a)], 
 colorof=[(a,1), (b,3), (c,2)], 
 colors=[1, 2, 3]
} 
\end{lstlisting}
\end{example}

In this step we also notice the consequences of file-based communications
with Clingo. In most situations, not all information in the answer
set must be presented to the end user. The current approach only lets
us filter the information after the solver has finished and printed
the complete answer set. A tighter integration with Clingo could overcome
this problem.

\section{Optimisations\label{sec:Optimisations}}

FMS contains a number of optimisations which are transformations of
Core expressions. Optimisations have the goal to make the expression
that needs to be translated simpler. In most cases this has a direct
correspondence to smaller expressions. Shorter expressions will lead
to a smaller translation, which leads to a smaller, but equivalent
SAT formula. This potentially results in exponential speedups. In
this section we give an overview of the implemented techniques.

\subsection{Stratifying Definitions}

Let-bindings are potentially very large, containing definitions for
a lot of different symbols. Translating mutually recursive definitions
leads to extra intermediate symbols in the translation process. Smaller,
but deeper nested bindings explicitate that there is no recursion
between symbols and result in better translations. We can use a topological
ordering of the bindings to transform the bindings. Tarjan's algorithm
is used to find this ordering. This has already been done in other
systems such as IDP3\cite{tplp/Jansen13}. As a side effect of this
topological ordering, definitions which are unused can be thrown away,
because the formula after the ``\lstinline!in!'' provides the starting
points for the stratification. If a definition is unconnected to the
formula, it is not included in the topological ordering.
\begin{example}
\begin{lstlisting}[breaklines=true]
let odd x := even (x-1) ; even x := if x = 0 then true else odd (x-1) ; c := 4 ; e := even c ; d := 8 in e.
\end{lstlisting}

can be stratified into:
\begin{lstlisting}[breaklines=true]
let odd x :=$\ldots$ ; even x :=$\ldots$ in 
	(let c := 4 in (let e := even c in e)).
\end{lstlisting}
\end{example}

\subsection{Inlining}

Let-bindings contain definitions for certain symbols. These symbols
occur in the inner expression of the let-binding. Sometimes it is
benificial to replace the defined symbol with its definition. It allows
for other optimisations and prevents the need to always define the
symbol. The process of replacing a symbol with its definition is called
inlining. Inlining is done in a lot of major compilers. The GHC compiler
for Haskell excels in this\cite{DBLP:journals/jfp/JonesM02} and we
copied some of their techniques into FMS. One simple observation is
that if a binding only has one single usage, it is always better to
optimise the binding for this usage, as we can remove the original
binding because it becomes unused.
\begin{example}
In the expression \lstinline!let y := 2*x in y+5!, the variable \lstinline!y!
can easily be inlined, so the new expression becomes: \lstinline!2*x+5!.
\end{example}

\begin{example}
In the expression \lstinline!let y := f x in y+y!, the variable \lstinline!y!
would not be inlined, as \lstinline!f x! would occur twice in the
resulting expression, which could duplicate work in further steps.
\end{example}

Another inlining rule is that bindings to constant integers or strings
should always be inlined, as handling those is generally not any more
complex than handling a variable, but allows for more specialised
optimisations. Sometimes the end user knows that inlining a function
would be beneficial but the compiler could not derive this. For this
reason FML allows for compiler directives which force the compiler
to inline (or not inline) certain functions.

\subsection{Boolean simplifications}

When handling complex nested boolean formulas, applying some boolean
rewrite rules can simplify them greatly. For this reason, pushing
of negations is implemented just like in IDP3\cite{DeCatPhd14}. This
technique uses the standard rewriting rules for propositional and
first order logic.
\begin{example}
The boolean expression \lstinline!not (or (not p) (not q))! is simplified
to the expression \lstinline!and p q!
\end{example}

\subsection{Constant folding and beta reductions}

Constant folding\cite{DBLP:books/cu/Appel1998c} is a common optimisation
technique for compilers. Whenever the arguments of builtin functions
like addition or multiplication are fully known, we can replace the
expression with its evaluation. This can be extended to more complex
expressions: if the ``if''-part of an if-then-else expression is
trivially true, the if-then-else can be replaced by just the then-part.

When the arguments of an anonymous function are known, we can specialise
the function body with the given arguments. This corresponds to the
concept of beta reduction in the lambda calculus. Using beta reduction
prevents the overhead of translating an extra lambda and can pair
up with other optimisations such as constant folding for more efficiency
gains. 
\begin{example}
\begin{sloppypar}

In the expression \lstinline!(\x -> x + 4) ((\x -> 5) a)!, the second
part \lstinline!((\x -> 5) a)! can be beta reduced to \lstinline!5!.
The resulting expression \lstinline!(\x -> x + 4) 5! can be beta
reduced to \lstinline!4 + 5! which can be reduced to \lstinline!9!
through constant folding. 

\end{sloppypar}
\end{example}

\subsection{Combining the optimisations}

All of the above optimisations can rewrite part of the expression
tree. The application of an optimisation possibly opens up possibilities
for other optimisations or sometimes, even the same optimisation.
To fully optimise an expression, all optimisations should be repeated
until a fixpoint on the expression is reached. 

\subsection{Future Optimisations}

All above optimisations are already present in the current system.
There are also a few optimisation techniques which have not been implemented
but could be interesting future expansions.

\subsubsection{Rewrite Rules}

Boolean simplifications are applications of rewrite rules which make
the expressions simpler. If the user writes new complex functions,
it can be beneficial that the end-user is able to write such rewrite
rules for his own functions as meta-information for his definitions.
This technique would be able to generalise the boolean simplifications
and make it applicable to more situations. Such rules are already
available in some programming languages such as Haskell.
\begin{example}
A rewriting rule like \lstinline!not (not x) => x! could explain
that double negations can be ignored.
\end{example}

\subsubsection{Ground with Bounds}

There are a lot of optimisations for grounding researched in the context
of the IDP3 system. While some are subsumed by new techniques which
are explained in this section, a lot of them are also applicable for
the FML system. Grounding with bounds\cite{aaai/WittocxMD08,jair/WittocxMD10}
is such a technique that proved to be signficant for the performance
of IDP3. Grounding with bounds uses symbolic reasoning to limit the
size of quantifications. It is our hypothesis that these techniques
generalize naturally to the higher order case of FML. This would allow
quantifications over implicit domains so the forall quantifier could
have type \lstinline!(a -> Bool) -> Bool! instead of type \lstinline!Set a -> (a -> Bool) -> Bool!.
\begin{example}
Ground with bounds could rewrite an unbounded quantification like
\lstinline*! (\x -> (5 < x & x > 10) => p x)* to a bounded expression
\lstinline*! {6..9} p* which can be handled by the algorithms which
are already in place.
\end{example}

\section{Dependencies and availability\label{sec:Implementation}}

The system is fully written in Kotlin, with a Gradle build system,
in conjunction with a set of libraries. The most important ones are
ANTLR4 for parsing, logback and slf4j for logging, JUnit for testing,
and klaxon for JSON communication.

The software and its source can be found online at \url{https://dtai.cs.kuleuven.be/krr/fms}.
There is support for testing the software online on the website itself
or you can download it yourself. The only external dependency is Clingo,
which can be downloaded from \url{https://potassco.org/clingo/}.

At the moment there is no IDE support for FML, but the development
of an implementation for the Language Server Protocol\cite{LSP} is
in the early stages of development. This is a generic framework which
editors can support to automate the integration with a new language.

\section{Conclusion}

This paper introduces FMS and its general internal workflow. As far
as the authors are aware this is the first attempt for using a higher-order
functional language as a modelling language. We introduced the Core
language and its syntactical extensions into the Full language. We
touched upon the different steps in the solving process and on the
optimisation techniques which are available in the system. We also
highlighted points where further work could significantly improve
the system. A thorough comparison with other systems such as SMT or
ASP is also an ambition for the close future. FMS is not a finished
product yet, but it introduces some interesting new concepts. 

\bibliographystyle{plain}
\bibliography{refs,krrlib}

\begin{thebibliography}{10}

\bibitem{DBLP:books/cu/Appel1998c}
Andrew~W. Appel.
\newblock {\em Modern Compiler Implementation in {C}}.
\newblock Cambridge University Press, 1998.

\bibitem{iclp/Blockeeletall12}
Hendrik Blockeel, Bart Bogaerts, Maurice Bruynooghe, Broes {De Cat}, Stef {De
  Pooter}, Marc Denecker, Anthony Labarre, Jan Ramon, and Sicco Verwer.
\newblock Modeling machine learning and data mining problems with
  {FO($\cdot$)}.
\newblock In Agostino Dovier and V{\'{\i}}tor Santos~Costa, editors, {\em
  Proceedings of the 28th International Conference on Logic Programming -
  Technical Communications (ICLP'12)}, volume~17 of {\em LIPIcs}, pages 14--25.
  Schloss Daghstuhl - Leibniz-Zentrum fuer Informatik, September 2012.

\bibitem{aspcore2}
Francesco Calimeri, Wolfgang Faber, Martin Gebser, Giovambattista Ianni, Roland
  Kaminski, Thomas Krennwallner, Nicola Leone, Francesco Ricca, and Torsten
  Schaub.
\newblock {ASP-Core-2} input language format.
\newblock Technical report, ASP Standardization Working Group, 2013.

\bibitem{iclp/DassevilleHBJD16}
Ingmar Dasseville, Matthias van~der Hallen, Bart Bogaerts, Gerda Janssens, and
  Marc Denecker.
\newblock A compositional typed higher-order logic with definitions.
\newblock In Manuel Carro, Andy King, Marina De~Vos, and Neda Saeedloei,
  editors, {\em ICLP'16}, volume~52 of {\em OASIcs}, pages 14.1--14.14. Schloss
  Dagstuhl, November 2016.

\bibitem{DeCatPhd14}
Broes {De Cat}.
\newblock {\em Separating Knowledge from Computation: An {FO($\cdot$)}
  Knowledge Base System and its Model Expansion Inference}.
\newblock PhD thesis, KU Leuven, Leuven, Belgium, May 2014.

\bibitem{DBLP:conf/nmr/IanniIPSC04}
Giovambattista Ianni, Giuseppe Ielpa, Adriana Pietramala, Maria~Carmela
  Santoro, and Francesco Calimeri.
\newblock Enhancing answer set programming with templates.
\newblock In James~P. Delgrande and Torsten Schaub, editors, {\em 10th
  International Workshop on Non-Monotonic Reasoning {(NMR} 2004), Whistler,
  Canada, June 6-8, 2004, Proceedings}, pages 233--239, 2004.

\bibitem{tplp/Jansen13}
Joachim Jansen, Albert Jorissen, and Gerda Janssens.
\newblock Compiling input$\ast$ {FO}($\cdot$) inductive definitions into tabled
  {P}rolog rules for {IDP3}.
\newblock {\em TPLP}, 13(4--5):691--704, 2013.

\bibitem{DBLP:journals/jfp/JonesM02}
Simon L.~Peyton Jones and Simon Marlow.
\newblock {Secrets of the Glasgow Haskell Compiler inliner}.
\newblock {\em J. Funct. Program.}, 12(4{\&}5):393--433, 2002.

\bibitem{LeuschelButler:FME03}
Michael Leuschel and Michael Butler.
\newblock Pro{B}: A model checker for {B}.
\newblock In Keijiro Araki, Stefania Gnesi, and Dino Mandrioli, editors, {\em
  FME 2003: Formal Methods}, LNCS 2805, pages 855--874. Springer-Verlag, 2003.

\bibitem{marlow2016desugaring}
Simon Marlow, Simon Peyton~Jones, Edward Kmett, and Andrey Mokhov.
\newblock Desugaring haskell's do-notation into applicative operations.
\newblock In {\em Proceedings of the 9th International Symposium on Haskell},
  pages 92--104. ACM, 2016.

\bibitem{LSP}
Microsoft.
\newblock {Language Server Protocol}, 2018.

\bibitem{conf/cp/NethercoteSBBDT07}
Nicholas Nethercote, Peter~J. Stuckey, Ralph Becket, Sebastian Brand,
  Gregory~J. Duck, and Guido Tack.
\newblock Minizinc: Towards a standard {CP} modelling language.
\newblock In C.~Bessiere, editor, {\em CP'07}, volume 4741 of {\em LNCS}, pages
  529--543. Springer, 2007.

\bibitem{parr2013definitive}
Terence Parr.
\newblock {\em The definitive ANTLR 4 reference}.
\newblock Pragmatic Bookshelf, 2013.

\bibitem{MIT/Pierce2002}
Benjamin~C. Pierce.
\newblock {\em Types and Programming Languages}.
\newblock MIT Press, Cambridge, MA, USA, 2002.

\bibitem{asp/2001}
Alessandro Provetti and Tran~Cao Son, editors.
\newblock {\em Answer Set Programming, Towards Efficient and Scalable Knowledge
  Representation and Reasoning, Proceedings of the 1st Intl. ASP'01 Workshop,
  Stanford, March 26-28, 2001}, 2001.

\bibitem{DBLP:journals/jfp/SchrijversSW09}
Tom Schrijvers, Peter~J. Stuckey, and Philip Wadler.
\newblock Monadic constraint programming.
\newblock {\em J. Funct. Program.}, 19(6):663--697, 2009.

\bibitem{GHC}
Simon Peyton-Jones Simon~Marlow.
\newblock {The Glasgow Haskell Compiler}, 2012.

\bibitem{DBLP:journals/mscs/Wadler92}
Philip Wadler.
\newblock Comprehending monads.
\newblock {\em Mathematical Structures in Computer Science}, 2(4):461--493,
  1992.

\bibitem{wazny2006type}
Jeremy~Richard Wazny.
\newblock {\em Type inference and type error diagnosis for Hindley/Milner with
  extensions}.
\newblock Citeseer, 2006.

\bibitem{aaai/WittocxMD08}
Johan Wittocx, Maarten Mari{\"e}n, and Marc Denecker.
\newblock Grounding with bounds.
\newblock In Dieter Fox and Carla~P. Gomes, editors, {\em AAAI}, pages
  572--577. AAAI Press, 2008.

\bibitem{jair/WittocxMD10}
Johan Wittocx, Maarten Mari{\"e}n, and Marc Denecker.
\newblock Grounding {FO} and {FO(ID)} with bounds.
\newblock {\em J. Artif. Intell. Res. (JAIR)}, 38:223--269, 2010.

\end{thebibliography}

\end{document}